\documentclass[10pt]{article}
\usepackage{amsthm,amsfonts,epsfig}
\usepackage{amsmath}
\usepackage[T1]{fontenc}
\usepackage{graphicx}
\usepackage{graphics}
\usepackage{placeins}
\usepackage{here}
\usepackage{float}
\usepackage{subfigure}
\usepackage{lmodern}
\usepackage{enumerate}
\usepackage{multirow}
\usepackage{setspace}
\usepackage[a4paper, text={17cm,22.5cm},top=1cm, left=2.5cm, right=2.5cm, includeheadfoot, headheight=1cm]{geometry}
\newcommand{\mat}[1]{\mbox{\boldmath{$#1$}}}
\begin{document}
  \title{Modeling Compositional Regression with uncorrelated and correlated errors: a Bayesian approach}
	\date{}

 \author{Taciana K. O. Shimizu,\,Francisco Louzada,\,Adriano K. Suzuki\,and\, Ricardo S. Ehlers\\
  {\small Department of Applied Mathematics \& Statistics, ICMC, University of São Paulo} \\
  {\small Av. Trabalhador Saocarlense, 400 CEP 13.566-590, São Carlos, SP, Brazil} \\
	{\small e-mail: \emph{taci\_kisaki@yahoo.com.br}, \emph{louzada@icmc.usp.br}, \emph{suzuki@icmc.usp.br} and \emph{ehlers@icmc.usp.br}}
}
\maketitle
	\begin{abstract}
  Compositional data consist of known compositions vectors whose components are positive and defined in the interval (0,1) representing proportions or fractions of a ``whole''. The sum of these components must be equal to one. Compositional data is present in different knowledge areas, as in geology, economy, medicine among many others. In this paper, we introduce a Bayesian analysis for compositional regression applying additive log-ratio (ALR) transformation and assuming uncorrelated and correlated errors. The Bayesian inference procedure based on Markov Chain Monte Carlo Methods (MCMC). The methodology is illustrated on an artificial and a real data set of volleyball.  
  
\end{abstract}
%


	\section{Introduction}
	
Compositional data are vectors of proportions specifying $G$ fractions as a whole. Such data often result when raw data are normalized or when data is obtained as proportions of a certain heterogeneous quantity.

By definition, a vector \mat{x} in the Simplex sample space is a composition, elements of this vector are components and the vectors set is compositional data \cite{AITCHISON1}. Therefore, for $\mat{x}=(x_{1}, x_{2},\ldots,x_{G})'$ to be a compositional vector, $x_{i}$ is non negative value, for $i = 1,\ldots,G$ and $x_{1}+x_{2}+\ldots+x_{G} = 1$.

The first model addopted for the analysis of the compositional data was the Dirichlet distribution. However, it requires that the correlation structure is wholly negative, a fact that is not observed for compositional data, in which some correlations are positive (see for example, Aitchison \cite{AITCHISON1}).

Aitchison and Shen \cite{AITCHISONSHEN} developed the logistic-Normal class of distributions transforming the $G$ component vector $\textbf{x}$ into a vector $\textbf{y}$ in $R^{G-1}$ and considering the Additive Log-Ratio (ALR) function. The use of Bayesian methods is a good alternative for the analysis of compositional data (see for example, Iyengar and Dey, \cite{IYENGAR,IYENGAR2}; or Tjelmeland and Lund, \cite{TJELMELAND}), especially considering Markov Chain Monte Carlo (MCMC) methods.
	
The main purpose of the paper is based on the Bayesian approach for the compositional regression model assuming correlated and uncorrelated normal errors. Usually, the data (attack, block, serve and opponent error) from this type of sport have compositional restrictions, i.e., they have dependence structure, being that standard existing methods to analyze multivariate data under the usual assumption of multivariate normal distribution (see for example, Johnson et. al \cite{JOHNSON}) are not appropriate to analyze them.

We consider a real data set that is related to the first and second rounds matches of Brazilian Men's Volleyball Super League 2011/2012 obtained from the website \cite{CBV}. The data concern the teams that played and won the games in such rounds; more specifically, the points of the team that won each game were defined as composition and the proportions of each composition are the volleyball skills, as attack, block, serves and errors of the opposite team.

The points of the winning team in each game were obtained by four components. We denoted $x_{1}$ the proportion of points in the attack, $x_{2}$ the proportion of points in the block, $x_{3}$ the proportion of points in the serve and $x_{4}$ the proportion of points in the errors of the opposite team.

The paper is organized as follows: Section 2 introduces the formulation of regression model applied through the Additive Log-Ratio (ALR) transformation; Section 3 reports a Bayesian analysis of the proposed model assuming correlated and uncorrelated Normal errors; Section 4 provides the results of the application to an artificial and a real data set related to the Brazilian Men's Volleyball Super League 2011/2012; finally, Section 5 ends the paper with some final remarks.

\vspace{0.3cm}
\section{Formulation of the Model}

We can consider $y_{ij}=H(x_{ij}/x_{iG})$, $i=1,...,n$ and $j=1,...,g$, being $H(\bullet)$ the chosen transformation function that assures resulting vector has real components, where $x_{ij}$ represents the $i$-th observation for the $j$-th component, such that $x_{i1}>0,\ldots,x_{iG}>0$ and $\sum_{j=1}^{G}x_{ij}=1$, for $i=1,...,n$.

The ALR transformation for the analysis of compositional data is given by
\begin{equation} \label{alr}
y_{ij}=H \left(\frac{x_{ij}}{x_{iG}}\right)=\mbox{log}\left(\frac{x_{ij}}{x_{iG}}\right).
\end{equation}

The regression model assuming ALR transformation for the response variables is given by
\begin{equation} \label{modeloreg}
 y_{ij}=\beta_{0j}+\beta_{1j}\mat{z_{i}}+\epsilon_{ij}, 
\end{equation}

\noindent
where $y_{ij}=(y_{i1},\ldots,y_{ig})$ is a vector $(1\times g)$ of response variables where $g=G-1$ and $G$ number of compositional data components; $\mat{z_{i}}$ is a vector of covariates associated to the $i$-th sample; $\beta_{0j}$ is a vector $(1\times g)$ intercepts; $\beta_{1j}$ is a vector ($p\times g$) regression coefficients and $\epsilon_{ij}$ are random errors, for $j=1,\ldots,G-1$ and $i=1,\ldots,n$.

\vspace{0.3cm}	
\section{Bayesian analysis considering ALR transformation}

This section presents a Bayesian analysis of the model (\ref{modeloreg}) with ALR transformation (\ref{alr}) applied to response variables and assuming multivariate Normal distribution for the correlated and uncorrelated errors.

\vspace{0.3cm}	
\subsection{Bayesian analysis considering ALR transformation assuming uncorrelated errors}

The points of the winning team in each game were obtained by four components. We denoted $x_{1}$ the proportion of points in the attack, $x_{2}$ the proportion of points in the block, $x_{3}$ the proportion of points in the serve and $x_{4}$ the proportion of points in the errors of the opposite team, i.e, they are the dependent variables defined for our study. On the other hand, we considered five independent variables (covariates): player who scored more points in the game belongs to the winning team $(z_{1})$, the winning team has won League at least once in the last twelve years $(z_{2})$, percentage of excellent reception of the winning team in the game $(z_{3})$ and percentage of excellent defense of the loser team in the game $(z_{4})$.

We assume an additive log-ratio (ALR) transformation given by $y_{i1}=\log(x_{i1}/x_{i4})$, $y_{i2}=\log(x_{i2}/x_{i4})$ and $y_{i3}=\log(x_{i3}/x_{i4})$. 

The regression model obtained to transformed data $y_{i1}$, $y_{i2}$ e $y_{i3}$ is given by
\begin{eqnarray} \label{modeltrans}
y_{i1}&=&\beta_{01}+\beta_{11}z_{i1}+\beta_{21}z_{i2}+\beta_{31}z_{i3}+\beta_{41}z_{i4}+\epsilon_{i1}, \nonumber\\
y_{i2}&=&\beta_{02}+\beta_{12}z_{i1}+\beta_{22}z_{i2}+\beta_{32}z_{i3}+\beta_{42}z_{i4}+\epsilon_{i2} \quad \mbox{and} \\
y_{i3}&=&\beta_{03}+\beta_{13}z_{i1}+\beta_{23}z_{i2}+\beta_{33}z_{i3}+\beta_{43}z_{i4}+\epsilon_{i3}, \nonumber
\end{eqnarray}

\noindent
where the covariates associated with the $i$-th game are described above, $y_{ij}$ represents the transformed proportion of the $j$-th component (attack, block, serve and errors of the opposite team) in the $i$-th game, $\beta_{0j}$ represents the mean of the points proportion in the $j$-th component related to component $x_{i4}$ (errors of the opposite team) for the team that did not win the Super League, $\beta_{1j}$, $\beta_{2j}$, $\beta_{3j}$, $\beta_{4j}$ indicate a possible covariate effect on the $i$-th game and $\epsilon_{ij}$ represents the error vector assuming independent random variables with a Normal distribution $N(\mathbf{0},\Sigma_{1})$, where $\mathbf{0}$ is a vector of zeros and $\Sigma_{1}$ variance-covariance matrix defined by
\begin{eqnarray*}
\Sigma_{1}&=\left(
        \begin{array} {ccc}
				\sigma_{1}^{2}&0&0 \nonumber\\
				0&\sigma_{2}^{2}&0 \\
				0&0&\sigma_{3}^{2} \nonumber
\end{array} \right),
\end{eqnarray*}

The likelihood function of parameters $\mat{\nu_{1}}=(\mat{\beta_{0}},\mat{\beta_{1}},\mat{\beta_{2}},\mat{\beta_{3}},\mat{\beta_{4}},\mat{\sigma}^{2})$ is given by
\begin{eqnarray} \label{vero1}
L(\mat{\nu_{1}}) \propto \prod_{j=1}^{3}(\sigma_{j}^{2})^{-n/2}\mbox{exp}\left(-\frac{1}{2\sigma_{j}^{2}}\sum_{i=1}^{n}\epsilon_{ij}^{2}\right),
\end{eqnarray}

\noindent
where
$\sum_{i=1}^{n}\epsilon_{ij}^{2}=\sum_{i=1}^{n}(y_{ij}-\beta_{0j}-\beta_{1j}z_{i1}-\beta_{2j}z_{i2}-\beta_{3j}z_{i3}-\beta_{4j}z_{i4})^{2}$, for $j=1,2,3$ and $i=1,\ldots,128$.

An alternative statistical approach for the analysis of compositional data is the use of Bayesian methods (see for example, Iyengar and Dey \cite{IYENGAR}; or Tjelmeland and Lund \cite{TJELMELAND}), especially considering Markov Chain Monte Carlo (MCMC) methods (see for example, Gelfand and Smith \cite{GELFAND}).
 
The Bayesian inference allows to associate previous knowledge of the parameters through a prior distribution. The Bayesian inference procedure for regression model (\ref{modeltrans}) considers proper prior distributions guaranteeing proper posterior distributions. Furthermore, it was ensuring non-informative prior distributions according to the fixed hyperparameters. Thus, we assume the following prior distributions for the parameters $\mat{\nu_{1}}$
\begin{align} \label{prior1}
  \beta_{0j}&\sim\mbox{N}(a_{0j},b_{0j}^{2}) \nonumber \\
	\beta_{lj}&\sim\mbox{N}(a_{lj},b_{lj}^{2})  \\
	\sigma_{j}^{2}&\sim\mbox{IG}(c_{j},d_{j}), \nonumber
\end{align}

\noindent
where IG(c,d) denotes an Inverse-Gamma distribution with mean $d/(c-1)$ and variance $d^{2}/[(c-1)^{2}(c-2)]$, for $c>2$; $a_{0j},b_{0j},a_{lj},b_{lj},c_{j}$ and $d_{j}$ are known hyperparameters, $j=1,\ldots,3$ and $l=1,\ldots,4$. 

All the parameters were assumed independent a priori.

Posterior summaries of interest for the model (\ref{modeltrans}) assuming prior distributions (\ref{prior1}) are given using simulated samples of the joint posterior distribution for $\mat{\nu_{1}}$ obtained using the Bayes formula, that is,
\begin{align*}
\pi(\mat{\beta_{0}},\mat{\beta_{1}},\mat{\beta_{2}},\mat{\beta_{3}},\mat{\beta_{4}},\mat{\sigma^{2}}|\mat{y}) &\propto \prod_{j=1}^{3}\mbox{exp}\left[-\frac{1}{2b_{0j}^{2}}(\beta_{0j}-a_{0j})^{2}\right]\times \prod_{j=1}^{3}\prod_{l=1}^{4}\mbox{exp}\left[-\frac{1}{2b_{lj}^{2}}(\beta_{lj}-a_{lj})^{2}\right] \nonumber \\
&\times \prod_{j=1}^{3} (\sigma_{j}^{2})^{-(c_{j}+1)}\mbox{exp}\left(-\frac{d_{j}}{\sigma_{j}^{2}}\right) \times \prod_{j=1}^{3} \left(\sigma_{j}^{2}\right)^{-n/2}\mbox{exp}\left(-\frac{1}{2\sigma_{j}^{2}}\sum_{i=1}^{n}\epsilon_{ij}^{2}\right).
\end{align*}

The conditional posterior densities using Gibbs sampling algorithm (Gelfand and Smith \cite{GELFAND}) for each parameter are given by,
\begin{align} \label{post1}
\textit{i)} \, \pi(\beta_{0j}|\mat{\beta_{1}},\mat{\beta_{2}},\mat{\beta_{3}},\mat{\beta_{4}},\mat{\sigma^{2}},\mat{y}) \sim N\left[\frac{a_{0j}\sigma_{j}^{2}+b_{0j}\displaystyle\sum_{i=1}^{n}\mu_{i}^{(j)}}{\sigma_{j}^{2}+nb_{0j}^{2}},\frac{b_{0j}^{2}\sigma_{j}^{2}}{\sigma_{j}^{2}+nb_{0j}^{2}}\right], 
\end{align} 

\begin{align} \label{post2}
\textit{ii)}  \, \pi(\beta_{lj}|\mat{\beta_{0}},\mat{\beta_{-l}},\mat{\sigma^{2}},\mat{y}) \sim N\left[\frac{a_{lj}\sigma_{j}^{2}+b_{lj}\displaystyle\sum_{i=1}^{n}z_{il}\theta_{i}^{(j)}}{\sigma_{j}^{2}+b_{lj}^{2}\displaystyle\sum_{i=1}^{n}z_{il}^{2}},\frac{b_{lj}^{2}\sigma_{j}^{2}}{\sigma_{j}^{2}+b_{lj}^{2}\displaystyle\sum_{i=1}^{n}z_{il}^{2}}\right] \, \mbox{and} 
\end{align}

\begin{align} \label{post3}
\textit{iii)} \, \pi(\sigma_{j}^{2}|\mat{\beta_{0}},\mat{\beta_{1}},\mat{\beta_{2}},\mat{\beta_{3}},\mat{\beta_{4}},\mat{y}) \sim IG\left[c_{j}+\frac{n}{2},d_{j}+\frac{1}{2}\displaystyle\sum_{i=1}^{n}\epsilon_{ij}^{2}\right],
\end{align} 

\noindent
where 

\begin{flushleft}
$\mu_{i}^{(j)}=y_{ij}-\sum_{l=1}^{4}\beta_{lj}z_{il}$ \\
$\theta_{i}^{(j)}=y_{ij}-\beta_{0j} \, \mbox{and}$ \\
$\epsilon_{ij}=y_{ij}-\beta_{0j}-\sum_{l=1}^{4}\beta_{lj}z_{il}, \, \mbox{for} \quad i=1,\ldots,n; \, j=1,2,3 \, \mbox{and} \, l=1,2,3,4.$
\end{flushleft}

For the estimation procedure we consider joint estimation where all the model parameters are estimated simultaneously in the MCMC algorithm. The conditional densities above (\ref{post1}), (\ref{post2}), (\ref{post3}) belong to any known parametric density family. Posterior summaries of interest for each model are simulated using standard MCMC methods through the Just Another Gibbs Sampler (JAGS) program (\cite{Plummer2003}). We used the rjags package (\cite{Plummer2011}) interacting with R software (\cite{R2011}). 

\vspace{0.3cm}	
\subsection{Bayesian analysis considering ALR transformation assuming correlated errors}

This section consider correlated errors for the model given in (\ref{modeltrans}) with multivariate Normal distribution, i.e., $\epsilon_{ij}$ represents the errors vector assumed to be dependent random variables with a multivariate normal distribution $N_{3}(\textbf{0},\Sigma_{2})$, where $\mat{0}$ is a vector of zeros and $\Sigma_{2}$ variance-covariance matrix is given by
\begin{eqnarray} \label{Sigma}
\Sigma_{2}&=\left(
        \begin{array} {ccc}
				\sigma_{1}^{2}	&	\rho_{12}\sigma_{1}\sigma_{2}	&	\rho_{13}\sigma_{1}\sigma_{3}	\\
        \rho_{12}\sigma_{1}\sigma_{2}	&	\sigma_{2}^{2}	&	\rho_{23}\sigma_{2}\sigma_{3}	\\
        \rho_{13}\sigma_{1}\sigma_{3}	&	\rho_{23}\sigma_{2}\sigma_{3}	&	\sigma_{3}^{2}	\\
\end{array} \right),
\end{eqnarray}

Considering the assumptions above, the likelihood function of parameters $\mathbf{\nu_{2}}=(\mat{\beta_{0}},\mat{\beta_{1}},\mat{\beta_{2}},\mat{\beta_{3}},\mat{\beta_{4}},\mat{\sigma^{2}},\mat{\rho})$ is given by
\begin{align*}
L(\mat{\nu_{2}}) & \propto |\Sigma_{2}|^{-n/2}\mbox{exp}\left\{-\frac{1}{2R}\left[\frac{(1-\rho_{23}^{2})}{\sigma_{1}^{2}}\sum_{i=1}^{n}\epsilon_{i1}^{2}+\frac{(1-\rho_{13}^{2})}{\sigma_{2}^{2}}\sum_{i=1}^{n}\epsilon_{i2}^{2}+\frac{(1-\rho_{12}^{2})}{\sigma_{3}^{2}}\sum_{i=1}^{n}\epsilon_{i3}^{2}\right]\right\} \\
& \times \mbox{exp}\left\{-\frac{1}{2R}\left[2R_{12}\sum_{i=1}^{n}\epsilon_{i1}\epsilon_{i2}+2R_{13}\sum_{i=1}^{n}\epsilon_{i1}\epsilon_{i3}+2R_{23}\sum_{i=1}^{n}\epsilon_{i2}\epsilon_{i3}\right]\right\},
\end{align*}

\noindent
where 
$R=2\rho_{12}\rho_{13}\rho_{23}-(\rho_{12}^{2}+\rho_{13}^{2}+\rho_{23}^{2})$; 
$R_{12}=(\rho_{13}\rho_{23}-\rho_{12})/\sigma_{1}\sigma_{2}$; \, $R_{13}=(\rho_{12}\rho_{23}-\rho_{13})/\sigma_{1}\sigma_{3}$; \, $R_{23}=(\rho_{12}\rho_{13}-\rho_{23})/\sigma_{2}\sigma_{3}$ \, \mbox{and} 
$\sum_{i=1}^{n}\epsilon_{ij}=\sum_{i=1}^{n}(y_{ij}-\beta_{0j}-\beta_{1j}z_{i1}-\beta_{2j}z_{i2}-\beta_{3j}z_{i3}-\beta_{4j}z_{i4})$, \mbox{for} $j=1,2,3$ \mbox{and} $i=1,\ldots,128$.

For the Bayesian analysis, we assume the same prior distributions (\ref{prior1}) for the $\beta_{0j}$, $\beta_{lj}$ and $\sigma_{j}$, $j=1,2,3$ and $l=1,\ldots,4$. The Uniform prior was considered for $\mat{\rho}=(\rho_{12},\rho_{13},\rho_{23})$ given by
\begin{align} \label{prior2}
\mat{\rho}&\sim\mbox{U}(-1,1)
\end{align}

All the parameters were assumed independent a priori.

Posterior summaries of interest for the model defined by (\ref{modeltrans}), but with correlated errors assuming priors distributions (\ref{prior1}) are given using simulated samples of the joint posterior distribution for $\mat{\nu_{2}}$ obtained using the Bayes formula, that is, 
\begin{align*}
\pi(\mat{\beta_{0}},\mat{\beta_{1}},\mat{\beta_{2}},\mat{\beta_{3}},\mat{\beta_{4}},\mat{\sigma^{2}},\mat{\rho}|\mat{y}) & \propto \prod_{j=1}^{3}\mbox{exp}\left[-\frac{1}{2b_{0j}^{2}}(\beta_{0j}-a_{0j})^{2}\right]\times \prod_{j=1}^{3}\prod_{l=1}^{4}\mbox{exp}\left[-\frac{1}{2b_{lj}^{2}}(\beta_{lj}-a_{lj})^{2}\right] \nonumber \\
& \times \prod_{j=1}^{3} (\sigma_{j}^{2})^{-(c_{j}+1)}\mbox{exp}\left(-\frac{d_{j}}{\sigma_{j}^{2}}\right) \nonumber \\
& \times |\Sigma_{2}|^{-n/2}\mbox{exp}\left\{-\frac{1}{2R}\left[\frac{(1-\rho_{23}^{2})}{\sigma_{1}^{2}}\sum_{i=1}^{n}\epsilon_{i1}^{2}+\frac{(1-\rho_{13}^{2})}{\sigma_{2}^{2}}\sum_{i=1}^{n}\epsilon_{i2}^{2}+\frac{(1-\rho_{12}^{2})}{\sigma_{3}^{2}}\sum_{i=1}^{n}\epsilon_{i3}^{2}\right]\right\} \nonumber \\
& \times \mbox{exp}\left\{-\frac{1}{2R}\left[2R_{12}\sum_{i=1}^{n}\epsilon_{i1}\epsilon_{i2}+2R_{13}\sum_{i=1}^{n}\epsilon_{i1}\epsilon_{i3}+2R_{23}\sum_{i=1}^{n}\epsilon_{i2}\epsilon_{i3}\right]\right\},
\end{align*}

\noindent
where 
$R=2\rho_{12}\rho_{13}\rho_{23}-(\rho_{12}^{2}+\rho_{13}^{2}+\rho_{23}^{2})$; 
$R_{12}=(\rho_{13}\rho_{23}-\rho_{12})/\sigma_{1}\sigma_{2}$; \, $R_{13}=(\rho_{12}\rho_{23}-\rho_{13})/\sigma_{1}\sigma_{3}$; \, $R_{23}=(\rho_{12}\rho_{13}-\rho_{23})/\sigma_{2}\sigma_{3}$ \, \mbox{and} 
$\sum_{i=1}^{n}\epsilon_{ij}=\sum_{i=1}^{n}(y_{ij}-\beta_{0j}-\beta_{1j}z_{i1}-\beta_{2j}z_{i2}-\beta_{3j}z_{i3}-\beta_{4j}z_{i4})$, \mbox{for} $j=1,2,3$ \mbox{and} $i=1,\ldots,128$.

The conditional posterior densities for each parameter are given by,
\begin{align} \label{post4}
\textit{i)} \, \pi(\beta_{0j}|\mat{\beta_{1}},\mat{\beta_{2}},\mat{\beta_{3}},\mat{\beta_{4}},\mat{\sigma^{2}},\mat{\rho},\mat{y}) & \propto \mbox{exp}\left[-\frac{1}{2b_{0j}^{2}}(\beta_{0j}-a_{0j})^{2}\right] \nonumber \\
& \times \mbox{exp}\left\{-\frac{1}{2R}\left[\frac{(1-\rho_{23}^{2})}{\sigma_{1}^{2}}\sum_{i=1}^{n}\epsilon_{i1}^{2}+\frac{(1-\rho_{13}^{2})}{\sigma_{2}^{2}}\sum_{i=1}^{n}\epsilon_{i2}^{2}+\frac{(1-\rho_{12}^{2})}{\sigma_{3}^{2}}\sum_{i=1}^{n}\epsilon_{i3}^{2}\right]\right\}  \nonumber \\
& \times \mbox{exp}\left\{-\frac{1}{2R}\left[2R_{12}\sum_{i=1}^{n}\epsilon_{i1}\epsilon_{i2}+2R_{13}\sum_{i=1}^{n}\epsilon_{i1}\epsilon_{i3}+2R_{23}\sum_{i=1}^{n}\epsilon_{i2}\epsilon_{i3}\right]\right\} , 
\end{align} 

\begin{align} \label{post5}
\textit{ii)}  \, \pi(\beta_{lj}|\mat{\beta_{0}},\mat{\beta_{-l}},\mat{\sigma^{2}},\mat{\rho},\mat{y}) & \propto \mbox{exp}\left[-\frac{1}{2b_{1j}^{2}}(\beta_{1j}-a_{1j})^{2}\right] \nonumber \\ 
& \times \mbox{exp}\left\{-\frac{1}{2R}\left[\frac{(1-\rho_{23}^{2})}{\sigma_{1}^{2}}\sum_{i=1}^{n}\epsilon_{i1}^{2}+\frac{(1-\rho_{13}^{2})}{\sigma_{2}^{2}}\sum_{i=1}^{n}\epsilon_{i2}^{2}+\frac{(1-\rho_{12}^{2})}{\sigma_{3}^{2}}\sum_{i=1}^{n}\epsilon_{i3}^{2}\right]\right\}  \nonumber \\ 
& \times \mbox{exp}\left\{-\frac{1}{2R}\left[2R_{12}\sum_{i=1}^{n}\epsilon_{i1}\epsilon_{i2}+2R_{13}\sum_{i=1}^{n}\epsilon_{i1}\epsilon_{i3}+2R_{23}\sum_{i=1}^{n}\epsilon_{i2}\epsilon_{i3}\right]\right\} , 
\end{align}

\begin{align} \label{post6}
\textit{iii)} \, \pi(\sigma_{j}^{2}|\mat{\beta_{0}},\mat{\beta_{1}},\mat{\beta_{2}},\mat{\beta_{3}},\mat{\beta_{4}},\mat{\rho},\mat{y}) & \propto \prod_{j=1}^{3} (\sigma_{j}^{2})^{-(c_{j}+1)}\mbox{exp}\left(-\frac{d_{j}}{\sigma_{j}^{2}}\right) \nonumber \\
& \times \mbox{exp}\left\{-\frac{1}{2R}\left[\frac{(1-\rho_{23}^{2})}{\sigma_{1}^{2}}\sum_{i=1}^{n}\epsilon_{i1}^{2}+\frac{(1-\rho_{13}^{2})}{\sigma_{2}^{2}}\sum_{i=1}^{n}\epsilon_{i2}^{2}+\frac{(1-\rho_{12}^{2})}{\sigma_{3}^{2}}\sum_{i=1}^{n}\epsilon_{i3}^{2}\right]\right\} \nonumber \\
& \times \mbox{exp}\left\{-\frac{1}{2R}\left[2R_{12}\sum_{i=1}^{n}\epsilon_{i1}\epsilon_{i2}+2R_{13}\sum_{i=1}^{n}\epsilon_{i1}\epsilon_{i3}+2R_{23}\sum_{i=1}^{n}\epsilon_{i2}\epsilon_{i3}\right]\right\}, 
\end{align} 

\begin{align} \label{post7}
\textit{iv)} \, \pi(\mat{\rho}|\mat{\beta_{0}},\mat{\beta_{1}},\mat{\beta_{2}},\mat{\beta_{3}},\mat{\beta_{4}},\mat{\sigma^{2}},\mat{y}) & \propto \mbox{exp}\left\{-\frac{1}{2R}\left[\frac{(1-\rho_{23}^{2})}{\sigma_{1}^{2}}\sum_{i=1}^{n}\epsilon_{i1}^{2}+\frac{(1-\rho_{13}^{2})}{\sigma_{2}^{2}}\sum_{i=1}^{n}\epsilon_{i2}^{2}+\frac{(1-\rho_{12}^{2})}{\sigma_{3}^{2}}\sum_{i=1}^{n}\epsilon_{i3}^{2}\right]\right\} \nonumber \\
& \times \mbox{exp}\left\{-\frac{1}{2R}\left[2R_{12}\sum_{i=1}^{n}\epsilon_{i1}\epsilon_{i2}+2R_{13}\sum_{i=1}^{n}\epsilon_{i1}\epsilon_{i3}+2R_{23}\sum_{i=1}^{n}\epsilon_{i2}\epsilon_{i3}\right]\right\}, 
\end{align} 

\noindent
where \\
$R=2\rho_{12}\rho_{13}\rho_{23}-(\rho_{12}^{2}+\rho_{13}^{2}+\rho_{23}^{2})$; \\
$R_{12}=(\rho_{13}\rho_{23}-\rho_{12})/\sigma_{1}\sigma_{2}$; \, $R_{13}=(\rho_{12}\rho_{23}-\rho_{13})/\sigma_{1}\sigma_{3}$; \, $R_{23}=(\rho_{12}\rho_{13}-\rho_{23})/\sigma_{2}\sigma_{3}$ \, \mbox{and} \\
$\sum_{i=1}^{n}\epsilon_{ij}=\sum_{i=1}^{n}(y_{ij}-\beta_{0j}-\beta_{1j}z_{i})$, \, \mbox{for} \, i=1,\ldots,n; \, j=1,2,3 \, \mbox{and} \, l=1,2,3,4. 

For the estimation procedure we consider joint estimation where all the model parameters are estimated simultaneously in the MCMC algorithm. The conditional densities above (\ref{post4}), (\ref{post5}), (\ref{post6}), (\ref{post7}) do not belong to any known parametric density family. Posterior summaries of interest for each model are simulated using standard MCMC methods through the Just Another Gibbs Sampler (JAGS) program (\cite{Plummer2003}). We used the rjags package (\cite{Plummer2011}) interacting with R software (\cite{R2011}).

\vspace{0.3cm}	
\section{Application}

This section reports a simulation study for the compositional data and illustrates an application of the proposed methodology through ALR transformation based on data related to proportions of the points of volleyball teams. 
\subsection{Simulation Study}

A simulation study was conducted to illustrate the proposed methodology. The artificial data were generated randomly from multivarite Normal distribution with $\mat{\mu}=(0.6,-1,-1.9)$ and $\mat{\sigma^{2}}= (0.06,0.2,0.3)$. We considered one dichotomized covariate, namely $z_{i1}$ (player who scored in the $i$-th game belongs to the winning team) generated through $z_{1} \sim$ Bernoulli $(0.8)$ and $z_{i2}$ (percentage of excellent reception of the winning team in the game) generated through $z_{2} \sim$ Normal $(0.5,0.1)$. 

The simulation study was based on 1000 samples generated for each case mentioned above. Sample sizes $n=70, 100$ and $150$ (number of volleyball games) were chosen and the parameters values were fixed as $\beta_{01}=0.5,\beta_{02}=-1,\beta_{03}=-2$, $\beta_{11}=\beta_{12}=\beta_{13}=0.1$, $\beta_{21}=\beta_{22}=\beta_{23}=0.1$ and $\sigma_{1}=\sigma_{2}=\sigma_{3}=1$.  We used the rjags package (\cite{Plummer2011}) interacting with R software (\cite{R2011}).

Table \ref{tabsims} shows the simulation results, i.e, mean, standard deviation (SD) and coverage probability (CP). The CP was stable and close to the nominal coverage.

\begin{table}[h] 
\centering{\caption{\small Simulation Data. Summary of the posterior distributions for the models parameters assuming uncorrelated and correlated errors.} 
\vspace*{0.1cm}
\scriptsize
\begin{tabular}{c|cccl|cccl}
\hline	
\multirow{3}{*}{Sample} & \multicolumn{4}{c|}{Model (\ref{modeltrans}) assuming} & \multicolumn{4}{c}{Model (\ref{modeltrans}) assuming} \\
\multirow{3}{*}{Size} & \multicolumn{4}{c|}{uncorrelated errors} & \multicolumn{4}{c}{correlated errors} \\ \cline{2-9}
& Parameter &	Mean	&	SD	&	CP & Parameter &	Mean & SD	&	CP  \\
\hline
	&	$\beta_{01}$	&	0.6009	&	0.1647	&	0.899	&	$\beta_{01}$	&	0.6011	&	0.1644	&	0.896	\\
	&	$\beta_{02}$	&	-0.9970	&	0.2985	&	0.908	&	$\beta_{02}$	&	-0.9960	&	0.2976	&	0.899	\\
	&	$\beta_{03}$	&	-1.9189	&	0.3575	&	0.917	&	$\beta_{03}$	&	-1.9164	&	0.3613	&	0.902	\\
	&	$\beta_{11}$	&	0.0023	&	0.0758	&	0.631	&	$\beta_{11}$	&	0.0024	&	0.0759	&	0.626	\\
	&	$\beta_{12}$	&	-0.0003	&	0.1362	&	0.824	&	$\beta_{12}$	&	0.0016	&	0.1387	&	0.806	\\
	&	$\beta_{13}$	&	0.0118	&	0.1608	&	0.865	&	$\beta_{13}$	&	0.0131	&	0.1619	&	0.868	\\
n=70	&	$\beta_{21}$	&	-0.0062	&	0.2996	&	0.881	&	$\beta_{21}$	&	-0.0064	&	0.2997	&	0.879	\\
	&	$\beta_{22}$	&	0.0025	&	0.5595	&	0.898	&	$\beta_{22}$	&	-0.0039	&	0.5572	&	0.895	\\
	&	$\beta_{23}$	&	0.0165	&	0.6537	&	0.900	&	$\beta_{23}$	&	0.0097	&	0.6637	&	0.907	\\
	&	$\sigma_{1}$	&	0.0622	&	0.0108	&	0.897	&	$\sigma_{1}$	&	0.0613	&	0.0107	&	0.898	\\
	&	$\sigma_{2}$	&	0.2055	&	0.0360	&	0.898	&	$\sigma_{2}$	&	0.2032	&	0.0350	&	0.895	\\
	&	$\sigma_{3}$	&	0.3068	&	0.0527	&	0.896	&	$\sigma_{3}$	&	0.3035	&	0.0536	&	0.886	\\
	&		&		&		&		&	$\rho_{12}$	&	0.4519	&	0.0986	&	0.883	\\
	&		&		&		&		&	$\rho_{13}$	&	0.3695	&	0.1066	&	0.882	\\
	&		&		&		&		&	$\rho_{23}$	&	0.2021	&	0.1204	&	0.879	\\
\hline																	
	&	$\beta_{01}$	&	0.6040	&	0.1401	&	0.884	&	$\beta_{01}$	&	0.6042	&	0.1398	&	0.884	\\
	&	$\beta_{02}$	&	-1.0016	&	0.2512	&	0.893	&	$\beta_{02}$	&	-0.9979	&	0.2519	&	0.897	\\
	&	$\beta_{03}$	&	-1.8968	&	0.3056	&	0.890	&	$\beta_{03}$	&	-1.8939	&	0.3106	&	0.898	\\
	&	$\beta_{11}$	&	-0.0024	&	0.0635	&	0.502	&	$\beta_{11}$	&	-0.0024	&	0.0635	&	0.496	\\
	&	$\beta_{12}$	&	0.000	&	0.1092	&	0.786	&	$\beta_{12}$	&	-0.0020	&	0.1124	&	0.767	\\
	&	$\beta_{13}$	&	-0.0030	&	0.1369	&	0.829	&	$\beta_{13}$	&	-0.0047	&	0.1382	&	0.823	\\
n=100	&	$\beta_{21}$	&	-0.0063	&	0.2585	&	0.856	&	$\beta_{21}$	&	-0.0067	&	0.2581	&	0.853	\\
	&	$\beta_{22}$	&	0.0036	&	0.4598	&	0.897	&	$\beta_{22}$	&	-0.0025	&	0.4585	&	0.879	\\
	&	$\beta_{23}$	&	-0.0093	&	0.5626	&	0.883	&	$\beta_{23}$	&	-0.0135	&	0.5750	&	0.882	\\
	&	$\sigma_{1}$	&	0.0610	&	0.0089	&	0.897	&	$\sigma_{1}$	&	0.0604	&	0.0088	&	0.890	\\
	&	$\sigma_{2}$	&	0.2031	&	0.0287	&	0.900	&	$\sigma_{2}$	&	0.2013	&	0.0287	&	0.905	\\
	&	$\sigma_{3}$	&	0.3068	&	0.0436	&	0.900	&	$\sigma_{3}$	&	0.3034	&	0.0425	&	0.906	\\
	&		&		&		&		&	$\rho_{12}$	&	0.4524	&	0.0803	&	0.889	\\
	&		&		&		&		&	$\rho_{13}$	&	0.3667	&	0.0846	&	0.900	\\
	&		&		&		&		&	$\rho_{23}$	&	0.2053	&	0.0950	&	0.907	\\
\hline																	
	&	$\beta_{01}$	&	0.5995	&	0.1143	&	0.894	&	$\beta_{01}$	&	0.5995	&	0.1144	&	0.893	\\
	&	$\beta_{02}$	&	-0.9865	&	0.2002	&	0.889	&	$\beta_{02}$	&	-0.9885	&	0.2024	&	0.897	\\
	&	$\beta_{03}$	&	-1.8908	&	0.2495	&	0.897	&	$\beta_{03}$	&	-1.8910	&	0.2456	&	0.897	\\
	&	$\beta_{11}$	&	0.0000	&	0.0526	&	0.371	&	$\beta_{11}$	&	0.0000	&	0.0525	&	0.367	\\
	&	$\beta_{12}$	&	-0.0062	&	0.0939	&	0.686	&	$\beta_{12}$	&	-0.0056	&	0.0917	&	0.703	\\
	&	$\beta_{13}$	&	0.0019	&	0.1128	&	0.780	&	$\beta_{13}$	&	0.0015	&	0.1113	&	0.787	\\
n=150	&	$\beta_{21}$	&	0.0023	&	0.2050	&	0.847	&	$\beta_{21}$	&	0.0023	&	0.2050	&	0.849	\\
	&	$\beta_{22}$	&	-0.0172	&	0.3649	&	0.883	&	$\beta_{22}$	&	-0.0131	&	0.3671	&	0.886	\\
	&	$\beta_{23}$	&	-0.0195	&	0.4545	&	0.880	&	$\beta_{23}$	&	-0.0177	&	0.4485	&	0.887	\\
	&	$\sigma_{1}$	&	0.0606	&	0.0069	&	0.910	&	$\sigma_{1}$	&	0.0602	&	0.0068	&	0.904	\\
	&	$\sigma_{2}$	&	0.2027	&	0.0235	&	0.916	&	$\sigma_{2}$	&	0.2012	&	0.0239	&	0.890	\\
	&	$\sigma_{3}$	&	0.3029	&	0.0362	&	0.898	&	$\sigma_{3}$	&	0.3010	&	0.0355	&	0.888	\\
	&		&		&		&		&	$\rho_{12}$	&	0.4523	&	0.0637	&	0.909	\\
	&		&		&		&		&	$\rho_{13}$	&	0.3701	&	0.0688	&	0.903	\\
	&		&		&		&		&	$\rho_{23}$	&	0.1995	&	0.0783	&	0.897	\\
\hline
\end{tabular}
\label{tabsims}}
\end{table}

Table \ref{tabcrit} shows the Bayesian criteria for the model assuming uncorrelated and correlated errors. The model assuming correlated errors is better when compared to the other model in all considered criteria.

\begin{table}[h] 
\centering{\caption{\small Simulation Data. Bayesian Criteria.} 
\vspace*{0.3cm}
\begin{tabular}{l|l|ccc}
\hline	
Sample & \multirow{2}{*}{Model} & \multicolumn{3}{c}{Bayesian criteria} \\ \cline{3-5}
Size & & EAIC & EBIC & DIC \\
\hline
\multirow{2}{*}{$n=70$}	&	Uncorrelated errors	&	215.11	&	242.09	&	203.78	\\	
	&	Correlated errors	&	-187.65	&	-160.67	&	-196.45	\\	\hline
\multirow{2}{*}{$n=100$}	&	Uncorrelated errors	&	301.02	&	332.28	&	289.49	\\	
	&	Correlated errors	&	-278.12	&	-246.85	&	-286.98	\\	\hline
\multirow{2}{*}{$n=150$}	&	Uncorrelated errors	&	444.69	&	480.82	&	432.99	\\	
	&	Correlated errors	&	-428.94	&	-392.82	&	-437.86	\\	
\hline
\end{tabular}
\label{tabcrit}}
\end{table}

\subsection{Real data application}

In this section, we consider a Bayesian analysis of the real data set presented in the Appendix (Table A) to illustrate an application of the proposed methodology, in particular, data related to proportions of the points volleyball teams. We apply the compositional data methodology to this set considering as components proportions the winning team points in 128 games of Brazilian Men's Volleyball Super League 2011/2012. This study was based on the four components: attack ($x_{i1}$), block ($x_{i2}$), serve ($x_{i3}$) and errors of the opposite team ($x_{i4}$), for $i=1,\ldots,128$.

The proposed model in (\ref{modeltrans}) and the following independent proper prior distributions (\ref{prior1}) were considered: $\beta_{0j} \sim N(0,1000)$, $\beta_{lj} \sim N(0,1000)$, $\sigma_{j}^{2} \sim IG(0.1,100)$, where $l=1,2,3,4$ and $j=1,2,3$. For proposed regression model with correlated errors (\ref{Sigma}), we considered the same independent proper prior distributions for $\beta_{0j}$, $\beta_{lj}$ and $\sigma_{j}^{2}$, for $l=1,2,3,4$ and $j=1,2,3$.  
It was simulated 100.000 Gibbs samples using the rjags package (\cite{Plummer2011}) interacting with R software (\cite{R2011}), in which the first 10.000 simulated samples were discarded to eliminate the effects of the initial values and we considered every 20th sample among the 90.000 Gibbs samples. The convergence was verified through Gelman-Rubin diagnostic. It shows values very close to 1 indicating convergence of the simulation algorithm.

According to Carlin and Louis (\cite{CARLIN}), the most basic tool for investigating model uncertainty is the \textit{sensitivity analysis}, that is, making reasonable modifications to the assumption, recomputing the posterior quantities of interest and seeing if they have changed in a way that has practical impact on interpretations. Thus, we checked the sensitivity analysis for different choices of prior parameters ($\beta_{0j}$, $\beta_{lj}$ and $\sigma_{j}^{2}$, for $l=1,2,3,4$) by changing only on parameter at a time and keeping all other parameters constant to their default values. We observe that posterior summaries of the parameters do not present considerable difference and not affect the results.

Table \ref{table1} shows the posterior summaries for the parameters of the model (\ref{modeltrans}) assuming uncorrelated and correlated errors based on these 4.500 final simulated Gibbs samples.

Note that, the estimated posterior means and standard deviations present similarity values for the both models (uncorrelated and correlated errors). 

Table \ref{table1} shows the posterior summaries for the parameters of the model (\ref{modeltrans}) assuming uncorrelated and correlated errors based on these 9.000 final simulated Gibbs samples. The convergence was verified through Gelman-Rubin diagnostic. It showed values very close to 1 indicating convergence of the simulation algorithm. Note that there is significant difference regarding to the proportions attack, block and serve points indicating by the estimated $\beta_{11}$, $\beta_{31}$, $\beta_{42}$ and $\beta_{43}$ for both models (uncorrelated and correlated errors), i.e., the player who scored in the game belongs to the winning team, percentage of excellent reception of the winning team and percentage of excellent defense of the loser team help it in these skills. Moreover, the estimated posterior means and standard deviations present similarity values for the both models (uncorrelated and correlated errors). We also observe that more parameters were significant in the correlated model than uncorrelated model, i.e., $\beta_{12},\beta_{13},\beta_{21},\beta_{22},\beta_{32}$ and $\beta_{41}$.

\begin{table}[h] 
\centering{\caption{\small Summary of the posterior distributions for the models parameters assuming uncorrelated and correlated errors.} 
\scriptsize
\vspace*{0.3cm}
\begin{tabular}{cccl|cccl}
\hline	
\multicolumn{4}{c|}{Model (\ref{modeltrans}) assuming} & \multicolumn{4}{c}{Model (\ref{modeltrans}) assuming} \\
\multicolumn{4}{c|}{uncorrelated errors} & \multicolumn{4}{c}{correlated errors} \\
\hline
\multirow{2}{*}{Parameter} &	\multirow{2}{*}{Mean}	&	Standard	&	Credibility & \multirow{2}{*}{Parameter} &	\multirow{2}{*}{Mean}	&	Standard	&	Credibility  \\
 & & Deviation & Interval (90\%) & & & Deviation & Interval (90\%) \\
\hline
$\beta_{01}$	&	0.5570	&	0.1918	&	(0.2406; 0.8693)	&	$\beta_{01}$	&	0.5402	&	0.1340	&	(0.3172; 0.7575)	\\
$\beta_{02}$	&	-1.9765	&	0.3457	&	(-2.5470; -1.4050)	&	$\beta_{02}$	&	-1.9646	&	0.2373	&	(-2.3511; -1.5750)	\\
$\beta_{03}$	&	-0.9372	&	0.4806	&	(-1.7273; -0.1537)	&	$\beta_{03}$	&	-0.9369	&	0.3362	&	(-1.4786; -0.3802)	\\
$\beta_{11}$	&	0.1729	&	0.0439	&	(0.0994; 0.2438)	&	$\beta_{11}$	&	0.1739	&	0.0302	&	(0.1248; 0.2243)	\\
$\beta_{12}$	&	0.1434	&	0.0787	&	(0.0149; 0.2741)	&	$\beta_{12}$	&	0.1422	&	0.0539	&	(0.0544; 0.2308)	\\
$\beta_{13}$	&	0.1896	&	0.1103	&	(0.0100; 0.3692)	&	$\beta_{13}$	&	0.1904	&	0.0763	&	(0.0640; 0.3157)	\\
$\beta_{21}$	&	-0.0719	&	0.0418	&	(-0.1414; -0.0032)	&	$\beta_{21}$	&	-0.0710	&	0.0288	&	(-0.1187; -0.0243)	\\
$\beta_{22}$	&	-0.1533	&	0.0748	&	(-0.2757; -0.0296)	&	$\beta_{22}$	&	-0.1540	&	0.0518	&	(-0.2393; -0.0682)	\\
$\beta_{23}$	&	-0.0269	&	0.1034	&	(-0.1973; 0.1431)	&	$\beta_{23}$	&	-0.0266	&	0.0721	&	(-0.1451; 0.0903)	\\
$\beta_{31}$	&	0.4273	&	0.1740	&	(0.1409; 0.7106)	&	$\beta_{31}$	&	0.4317	&	0.1235	&	(0.2305; 0.6345)	\\
$\beta_{32}$	&	0.5267	&	0.3124	&	(0.0052; 1.0393)	&	$\beta_{32}$	&	0.5156	&	0.2156	&	(0.1631; 0.8730)	\\
$\beta_{33}$	&	-0.2667	&	0.4356	&	(-0.9798; 0.4562)	&	$\beta_{33}$	&	-0.2705	&	0.3049	&	(-0.7736; 0.2224)	\\
$\beta_{41}$	&	-0.5559	&	0.2818	&	(-1.0130; -0.0973)	&	$\beta_{41}$	&	-0.5284	&	0.1945	&	(-0.8450; -0.2098)	\\
$\beta_{42}$	&	1.2419	&	0.5086	&	(0.4003; 2.0773)	&	$\beta_{42}$	&	1.2339	&	0.3470	&	(0.6665; 1.8018)	\\
$\beta_{43}$	&	-1.9218	&	0.7063	&	(-3.0970; -0.7714)	&	$\beta_{43}$	&	-1.9212	&	0.4873	&	(-2.7199; -1.1283)	\\
$\sigma_{1}$	&	0.0515	&	0.0067	&	(0.0415; 0.0635)	&	$\sigma_{1}$	&	0.0242	&	0.0022	&	(0.0209; 0.0280)	\\
$\sigma_{2}$	&	0.1634	&	0.0213	&	(0.1317; 0.2011)	&	$\sigma_{2}$	&	0.0775	&	0.0070	&	(0.0667; 0.0895)	\\
$\sigma_{3}$	&	0.3172	&	0.0411	&	(0.2560; 0.3897)	&	$\sigma_{3}$	&	0.1524	&	0.0137	&	(0.1310; 0.1763)	\\
				&	&	&	&	$\rho_{12}$	&	0.1296	&	0.0399	&	(0.0632; 0.1946)	\\
				&	&	&	&	$\rho_{13}$	&	0.0793	&	0.0396	&	(0.0137; 0.1443)	\\
				&	&	&	&	$\rho_{23}$	&	0.0514	&	0.0397	&	(-0.0145; 0.1167)	\\
\hline
\end{tabular}
\label{table1}}
\end{table}

Table \ref{table2} presents the Bayesian model selection criteria expected Akaike information criterion (\textit{EAIC}), expected Bayesian information criterion (\textit{EBIC}), deviance information criterion (\textit{DIC}) and summary statistics of the $\mbox{CPO}_{i}'s$ $(LPML=\sum_{i=1}^{n}\mbox{log}(\widehat{CPO}))$. These results are suggesting that fitted regression model assuming correlated errors is the best choice (lower values EAIC, EBIC and DIC).

\begin{table}[!h] 
\centering{\caption{Bayesian Criteria for the models parameters assuming uncorrelated and correlated errors.} 
\vspace*{0.3cm}
\begin{tabular}{lcccc}
\hline	
\multirow{2}{*}{Model} & \multicolumn{4}{c}{Bayesian Criteria} \\ \cline{2-5} 
 & EAIC & EBIC & DIC & LPML \\
\hline
Uncorrelated errors	&	329.924	&	363.959	&	325.284	&	152.962	\\
Correlated errors	&	-599.408	&	-539.846	&	-619.624	& -320.704	\\
\hline
\end{tabular}
\label{table2}}
\end{table}

\vspace{0.3cm}	
\section{Concluding Remarks}

In this paper, we present a Bayesian analysis for compositional regression model considering ALR transformation and assuming uncorrelated and correlated errors. The inferencial procedure for the parameters based on MCMC methods. We have illustrated the proposed methodology considering a real data set from percentages of winning volleyball team's points, in which it was considered multivariate data structure. A comparation study of models was carried out through model selection procedure based on a statistical criteria. Thus, the results indicate that the compositional regression model with correlated errors outperforms the model with uncorrelated errors, besides pointing out the advantage of considering the natural multivariate structure of the data.

\nocite*{}

\newpage
\section*{Appendix}
\subsection*{Table A. Matches of Brazilian Men's Volleyball Super League 2011/2012.}

\begin{table}[ht!]
\centering 
\resizebox{!}{10cm}{ 
\begin{tabular}{c|cccccccc} 
\hline	 
Matches	&	\% attack	&	\% block	&	\% serve	&	\% errors	&	$z_{1}$ &	$z_{2}$	&	$z_{3}$	&	$z_{4}$	\\
\hline	 
1	&	48.00	&	12.00	&	2.67	&	37.33	&	0	&	1	&	0.6591	&	0.3958	\\
2	&	53.06	&	14.29	&	7.14	&	25.51	&	1	&	1	&	0.3467	&	0.4394	\\
3	&	44.00	&	13.33	&	8.00	&	34.67	&	1	&	0	&	0.4500	&	0.3571	\\
4	&	52.63	&	14.74	&	7.37	&	25.26	&	1	&	0	&	0.4605	&	0.6000	\\
5	&	56.00	&	8.00	&	5.33	&	30.67	&	0	&	1	&	0.7708	&	0.5192	\\
6	&	65.63	&	10.16	&	2.34	&	21.88	&	1	&	0	&	0.5247	&	0.4835	\\
7	&	54.67	&	14.67	&	2.67	&	28.00	&	1	&	1	&	0.6735	&	0.4746	\\
8	&	50.00	&	12.50	&	9.62	&	27.88	&	0	&	1	&	0.5000	&	0.5075	\\
9	&	52.58	&	15.46	&	3.09	&	28.87	&	1	&	0	&	0.6452	&	0.5500	\\
10	&	57.33	&	13.33	&	4.00	&	25.33	&	1	&	1	&	0.5926	&	0.5091	\\
11	&	56.60	&	11.32	&	6.60	&	25.47	&	1	&	0	&	0.6447	&	0.4605	\\
12	&	49.33	&	9.33	&	14.67	&	26.67	&	1	&	1	&	0.5555	&	0.4082	\\
13	&	56.67	&	8.33	&	6.67	&	28.33	&	1	&	0	&	0.3936	&	0.4857	\\
14	&	66.67	&	4.00	&	1.33	&	28.00	&	1	&	1	&	0.7924	&	0.4038	\\
15	&	56.70	&	10.31	&	3.09	&	29.90	&	1	&	0	&	0.5867	&	0.4677	\\
16	&	62.67	&	6.67	&	4.00	&	26.67	&	1	&	1	&	0.4561	&	0.3958	\\
17	&	49.35	&	5.19	&	3.90	&	41.56	&	0	&	1	&	0.2453	&	0.4510	\\
18	&	48.00	&	13.33	&	6.67	&	32.00	&	1	&	1	&	0.5000	&	0.4808	\\
19	&	56.12	&	14.29	&	4.08	&	25.51	&	1	&	0	&	0.5507	&	0.4328	\\
20	&	44.00	&	9.33	&	5.33	&	41.33	&	1	&	0	&	0.5641	&	0.5172	\\
21	&	54.95	&	10.81	&	5.41	&	28.83	&	1	&	1	&	0.6173	&	0.2817	\\
22	&	59.34	&	15.38	&	3.30	&	21.98	&	0	&	0	&	0.5952	&	0.5889	\\
23	&	51.52	&	12.12	&	1.01	&	35.35	&	1	&	0	&	0.3194	&	0.5614	\\
24	&	57.69	&	14.10	&	5.13	&	23.08	&	1	&	0	&	0.6250	&	0.5000	\\
25	&	53.33	&	14.67	&	2.67	&	29.33	&	1	&	0	&	0.4800	&	0.7017	\\
26	&	55.45	&	13.86	&	5.94	&	24.75	&	0	&	1	&	0.8133	&	0.4146	\\
27	&	48.04	&	10.78	&	3.92	&	37.25	&	0	&	0	&	0.4722	&	0.5077	\\
28	&	48.00	&	17.33	&	9.33	&	25.33	&	1	&	0	&	0.6792	&	0.3846	\\
29	&	57.33	&	8.00	&	6.67	&	28.00	&	1	&	1	&	0.5217	&	0.4186	\\
30	&	53.78	&	8.40	&	5.04	&	32.77	&	1	&	1	&	0.3736	&	0.5488	\\
31	&	44.74	&	10.53	&	7.02	&	37.72	&	0	&	1	&	0.6265	&	0.4810	\\
32	&	53.57	&	13.10	&	3.57	&	29.76	&	1	&	1	&	0.5538	&	0.5082	\\
33	&	59.41	&	5.94	&	6.93	&	27.72	&	1	&	1	&	0.5556	&	0.5429	\\
34	&	59.21	&	9.21	&	6.58	&	25.00	&	1	&	0	&	0.5185	&	0.4792	\\
35	&	51.04	&	13.54	&	6.25	&	29.17	&	1	&	0	&	0.5410	&	0.5484	\\
36	&	50.51	&	13.13	&	3.03	&	33.33	&	0	&	0	&	0.3472	&	0.5454	\\
37	&	55.34	&	16.50	&	4.85	&	23.30	&	1	&	0	&	0.6420	&	0.5000	\\
38	&	61.33	&	4.00	&	2.67	&	32.00	&	1	&	0	&	0.3542	&	0.4681	\\
39	&	48.98	&	17.35	&	2.04	&	31.63	&	0	&	1	&	0.5758	&	0.4844	\\
40	&	55.32	&	9.57	&	5.32	&	29.79	&	1	&	1	&	0.4306	&	0.4576	\\
41	&	56.25	&	7.29	&	5.21	&	31.25	&	1	&	1	&	0.5454	&	0.4627	\\
42	&	56.19	&	7.62	&	2.86	&	33.33	&	1	&	0	&	0.4096	&	0.5286	\\
43	&	51.72	&	8.62	&	6.03	&	33.62	&	0	&	1	&	0.4762	&	0.4444	\\
44	&	49.46	&	13.98	&	3.23	&	33.33	&	1	&	1	&	0.6618	&	0.5714	\\
45	&	47.27	&	11.82	&	5.45	&	35.45	&	0	&	0	&	0.2250	&	0.5857	\\
46	&	57.33	&	13.33	&	1.33	&	28.00	&	0	&	0	&	0.5800	&	0.4461	\\
47	&	56.84	&	14.74	&	3.16	&	25.26	&	1	&	0	&	0.3947	&	0.6173	\\
48	&	60.61	&	10.10	&	3.03	&	26.26	&	1	&	0	&	0.5479	&	0.4756	\\
49	&	60.18	&	10.62	&	3.54	&	25.66	&	1	&	0	&	0.5312	&	0.5714	\\
50	&	52.43	&	9.71	&	3.88	&	33.98	&	0	&	0	&	0.6265	&	0.5000	\\
51	&	50.67	&	13.33	&	4.00	&	32.00	&	0	&	0	&	0.6000	&	0.5769	\\
52	&	63.30	&	8.26	&	4.59	&	23.85	&	1	&	1	&	0.4831	&	0.5000	\\
53	&	54.46	&	4.95	&	2.97	&	37.62	&	0	&	1	&	0.6493	&	0.5441	\\
54	&	56.25	&	8.33	&	4.17	&	31.25	&	1	&	0	&	0.3485	&	0.5769	\\
55	&	69.89	&	5.38	&	5.38	&	19.35	&	1	&	0	&	0.6479	&	0.4219	\\
56	&	65.82	&	15.19	&	3.80	&	15.19	&	1	&	0	&	0.4062	&	0.2857	\\
57	&	57.89	&	5.26	&	11.84	&	25.00	&	1	&	1	&	0.6415	&	0.3559	\\
58	&	36.84	&	12.63	&	7.37	&	43.16	&	0	&	0	&	0.4286	&	0.6271	\\
59	&	50.00	&	14.42	&	1.92	&	33.65	&	0	&	0	&	0.5672	&	0.5699	\\
60	&	52.08	&	6.25	&	5.21	&	36.46	&	0	&	0	&	0.6000	&	0.4638	\\
61	&	53.33	&	13.33	&	5.33	&	28.00	&	1	&	0	&	0.4706	&	0.4091	\\
62	&	44.00	&	16.00	&	10.67	&	29.33	&	1	&	1	&	0.4750	&	0.5238	\\
63	&	58.67	&	10.67	&	8.00	&	22.67	&	1	&	1	&	0.5102	&	0.4375	\\
64	&	54.00	&	8.00	&	5.00	&	33.00	&	0	&	0	&	0.6765	&	0.4545	\\ 
\end{tabular} }
\label{data} 
\end{table} 

\newpage
\begin{table}[ht!]
\centering 
\resizebox{!}{10cm}{ 
\begin{tabular}{c|cccccccc}
Matches	&	\% attack	&	\% block	&	\% serve	&	\% errors	&	$z_{1}$ &	$z_{2}$	&	$z_{3}$	&	$z_{4}$ \\
\hline	
65	&	53.27	&	9.35	&	5.61	&	31.78	&	1	&	1	&	0.4884	&	0.5067	\\
66	&	51.09	&	6.52	&	4.35	&	38.04	&	1	&	1	&	0.5373	&	0.5775	\\
67	&	55.26	&	6.58	&	6.58	&	31.58	&	1	&	1	&	0.6316	&	0.3889	\\
68	&	55.10	&	12.24	&	3.06	&	29.59	&	1	&	0	&	0.6349	&	0.3333	\\
69	&	64.94	&	6.49	&	5.19	&	23.38	&	0	&	1	&	0.6949	&	0.3906	\\
70	&	57.33	&	10.67	&	10.67	&	21.33	&	1	&	1	&	0.5319	&	0.4286	\\
71	&	56.70	&	12.37	&	7.22	&	23.71	&	1	&	0	&	0.7917	&	0.4412	\\
72	&	44.74	&	9.21	&	7.89	&	38.16	&	0	&	0	&	0.3182	&	0.5849	\\
73	&	62.50	&	13.54	&	4.17	&	19.79	&	1	&	1	&	0.4937	&	0.5352	\\
74	&	56.52	&	5.22	&	6.09	&	32.17	&	1	&	1	&	0.4494	&	0.4937	\\
75	&	53.19	&	14.89	&	4.26	&	27.66	&	1	&	0	&	0.5362	&	0.4237	\\
76	&	57.33	&	9.33	&	6.67	&	26.67	&	1	&	1	&	0.6491	&	0.4118	\\
77	&	50.67	&	18.67	&	2.67	&	28.00	&	1	&	1	&	0.4783	&	0.4706	\\
78	&	53.95	&	10.53	&	6.58	&	28.95	&	1	&	0	&	0.5000	&	0.4898	\\
79	&	44.00	&	16.00	&	2.67	&	37.33	&	1	&	1	&	0.5217	&	0.5208	\\
80	&	48.00	&	5.33	&	6.67	&	40.00	&	0	&	1	&	0.5333	&	0.2222	\\
81	&	56.00	&	8.00	&	2.67	&	33.33	&	1	&	0	&	0.7857	&	0.4048	\\
82	&	57.84	&	11.76	&	2.94	&	27.45	&	0	&	0	&	0.6512	&	0.4931	\\
83	&	47.25	&	12.09	&	4.40	&	36.26	&	1	&	1	&	0.5781	&	0.5443	\\
84	&	59.81	&	11.21	&	3.74	&	25.23	&	1	&	0	&	0.3483	&	0.5000	\\
85	&	53.06	&	11.22	&	5.10	&	30.61	&	0	&	1	&	0.4127	&	0.4091	\\
86	&	54.55	&	7.27	&	4.55	&	33.64	&	1	&	0	&	0.3333	&	0.4789	\\
87	&	59.18	&	6.12	&	3.06	&	31.63	&	1	&	1	&	0.4923	&	0.4203	\\
88	&	46.88	&	13.54	&	3.13	&	36.46	&	0	&	1	&	0.6351	&	0.5600	\\
89	&	57.50	&	11.67	&	1.67	&	29.17	&	0	&	0	&	0.6210	&	0.4713	\\
90	&	48.00	&	20.00	&	8.00	&	24.00	&	1	&	0	&	0.2391	&	0.5532	\\
91	&	54.00	&	13.00	&	2.00	&	31.00	&	0	&	1	&	0.6667	&	0.4068	\\
92	&	54.67	&	6.67	&	8.00	&	30.67	&	1	&	1	&	0.5870	&	0.3696	\\
93	&	56.12	&	7.14	&	3.06	&	33.67	&	1	&	1	&	0.5405	&	0.4930	\\
94	&	56.00	&	8.00	&	8.00	&	28.00	&	0	&	1	&	0.6226	&	0.4528	\\
95	&	50.67	&	13.33	&	6.67	&	29.33	&	1	&	0	&	0.5116	&	0.4615	\\
96	&	52.00	&	12.00	&	1.33	&	34.67	&	1	&	1	&	0.4762	&	0.4210	\\
97	&	58.06	&	10.75	&	2.15	&	29.03	&	1	&	0	&	0.4265	&	0.4776	\\
98	&	45.33	&	6.67	&	2.67	&	45.33	&	0	&	0	&	0.5200	&	0.5102	\\
99	&	51.90	&	15.19	&	2.53	&	30.38	&	1	&	0	&	0.5893	&	0.6461	\\
100	&	52.78	&	15.74	&	1.85	&	29.63	&	0	&	1	&	0.4348	&	0.4857	\\
101	&	54.29	&	9.52	&	7.62	&	28.57	&	1	&	0	&	0.3537	&	0.5422	\\
102	&	50.67	&	14.67	&	5.33	&	29.33	&	1	&	0	&	0.4595	&	0.4324	\\
103	&	48.00	&	20.00	&	2.67	&	29.33	&	1	&	1	&	0.6190	&	0.4348	\\
104	&	62.03	&	6.33	&	2.53	&	29.11	&	1	&	1	&	0.7736	&	0.4490	\\
105	&	40.00	&	12.00	&	9.33	&	38.67	&	0	&	0	&	0.6190	&	0.5526	\\
106	&	59.09	&	8.18	&	4.55	&	28.18	&	1	&	1	&	0.5581	&	0.4634	\\
107	&	49.33	&	9.33	&	8.00	&	33.33	&	1	&	1	&	0.6667	&	0.5263	\\
108	&	58.25	&	6.80	&	3.88	&	31.07	&	0	&	0	&	0.5976	&	0.4225	\\
109	&	60.78	&	9.80	&	6.86	&	22.55	&	0	&	0	&	0.6265	&	0.4651	\\
110	&	55.77	&	13.46	&	1.92	&	28.85	&	0	&	0	&	0.3368	&	0.6234	\\
111	&	56.00	&	8.00	&	4.00	&	32.00	&	0	&	0	&	0.3774	&	0.4286	\\
112	&	55.88	&	10.78	&	5.88	&	27.45	&	1	&	0	&	0.5135	&	0.5555	\\
113	&	64.13	&	5.43	&	7.61	&	22.83	&	1	&	0	&	0.6329	&	0.3973	\\
114	&	46.67	&	8.00	&	9.33	&	36.00	&	1	&	1	&	0.3409	&	0.4722	\\
115	&	56.14	&	8.77	&	4.39	&	30.70	&	0	&	1	&	0.5679	&	0.5185	\\
116	&	49.00	&	9.00	&	6.00	&	36.00	&	0	&	1	&	0.5540	&	0.5614	\\
117	&	54.67	&	10.67	&	5.33	&	29.33	&	1	&	1	&	0.5778	&	0.4348	\\
118	&	48.65	&	18.02	&	4.50	&	28.83	&	0	&	1	&	0.6429	&	0.6364	\\
119	&	64.22	&	7.34	&	1.83	&	26.61	&	1	&	1	&	0.5432	&	0.4255	\\
120	&	56.58	&	7.89	&	11.84	&	23.68	&	0	&	1	&	0.5319	&	0.3725	\\
121	&	56.38	&	14.89	&	5.32	&	23.40	&	1	&	0	&	0.4571	&	0.4286	\\
122	&	48.45	&	15.46	&	5.15	&	30.93	&	1	&	0	&	0.5833	&	0.4918	\\
123	&	52.83	&	6.60	&	6.60	&	33.96	&	1	&	0	&	0.2625	&	0.4927	\\
124	&	60.00	&	5.33	&	9.33	&	25.33	&	1	&	0	&	0.5283	&	0.3750	\\
125	&	59.63	&	9.17	&	1.83	&	29.36	&	1	&	1	&	0.5952	&	0.5484	\\
126	&	54.67	&	10.67	&	1.33	&	33.33	&	0	&	1	&	0.5965	&	0.6349	\\
127	&	54.67	&	6.67	&	6.67	&	32.00	&	1	&	0	&	0.3696	&	0.4103	\\
128	&	46.91	&	6.17	&	4.94	&	41.98	&	0	&	1	&	0.6863	&	0.5366	\\
\hline
\end{tabular} }
\label{data}
\end{table}


\begin{thebibliography}{999}


\bibitem{ACHCAROBAGE}
{ACHCAR, J. A.; OBAGE, S. C. Uma abordagem Bayesiana para dados composicionais considerando erros correlacionados. {\it Revista de Matemática e Estatística}, 23(2), 95-107, 2005.}

\bibitem{AITCHISON1}
{AITCHISON, J. The statistical analysis of compositional data. {\it Journal of the Royal Statistical Society. Series B (Methodological)}, 44(2), 139-177, 1982.}

\bibitem{AITCHISON2}
{AITCHISON, J. {\it The statistical analysis of compositional data}. Chapman \& Hall, 1986.}

\bibitem{AITCHISONSHEN}
{AITCHISON, J.; SHEN, S. M.  Logistic-normal distributions: Some properties and uses. {\it Biometrika}, 67(2), 261-272, 1980.}

\bibitem{BOXCOX}
{BOX, G. E. P.; COX, D. R. An analysis of transformations. {\it Journal of the Royal Statistical Society. Series B (Methodological)}, 26(2), p. 211-252, 1964.}

\bibitem{CBV}
{Brazilian Volleyball Confederation (CBV). {\it Data set of Men's Volleyball Super League}. Avaliable at: http://www.cbv.com.br/v1/superliga-1112/m-tabela.asp. Accessed August 23, 2012.}

\bibitem{CAMPOS}
{CAMPOS, F. A. D.; STANGANÉLLI, L. C. R.; CAMPOS, L. C. B.; PASQUARELLI, B. N.; GÓMEZ, M. A. Performance indicators analysis at Brazilian and Italian women's volleyball leagues according to game location, game outcome, and set number. {\it Perceptual and motor skills}, 118(2), 347-361, 2014.}

\bibitem{CARLIN}
{CARLIN, B. P.; LOUIS, T. A. {\it Bayesian Methods for Data Analysis}. Chapman \& Hall/CRC Texts in Statistical Science, 3th. Edition, 2009.}

\bibitem{FIVB}
{International Volleyball Federation (FIVB). Avaliable at: http://www.fivb.org/en/volleyball/History.asp. Accessed August 20, 2014.}

\bibitem{GELFAND}
{GELFAND, A. E.; SMITH, A. F. M.  Sampling based approaches to calculating marginal densities. {\it Journal of the American Statistical Association}, 85(410), 398-409, 1990.}

\bibitem{IYENGAR}
{IYENGAR, M.; DEY, D. K. {\it Bayesian analysis of compositional data}. Department of Statistics, University of Connecticut, Storrs, CT 06269-3120, 1996.}

\bibitem{IYENGAR2}
{IYENGAR, M.; DEY, D. K. {\it Box-Cox transformations in Bayesian analysis of compositional data}. {\it Environmetrics}, 9(6), 657-671, 1998.}

\bibitem{JOHNSON}
{JOHNSON, R.; WICHERN, D. {\it Applied multivariate statistical analysis.} New Jersey: Prentice Hall, 1998.}

\bibitem{LUNN}
{LUNN, D.; SPIEGELHALTER, D.; THOMAS, A.; BEST, N. The BUGS project: Evolution, critique and future directions. \textit{Statistics in Medicine}, 28(25), 3049–3067, 2009.}

\bibitem{Plummer2003}
{PLUMMER, M. JAGS: A program for analysis {B}ayesian graphical models using {G}ibbs sampling. {\it Proceedings of the 3rd International Workshop on Distributed Statistical Computing (DSC, 2003)}. Vol. 124. Technische Universit at Wien, 2003.}

\bibitem{Plummer2011}
{PLUMMER, M. rjags: Bayesian graphical models using MCMC. r package version 3-3, 2011. }

\bibitem{R2011}
{R Development Core Team. R: A Language and Environment for Statistical Computing. R Foundation for Statistical Computing, Vienna, Austria, 2011.}

\bibitem{RAYENS1}
{RAYENS, W. S.; SRINIVASAN, C. Box-Cox transformations in the analysis of compositional data. {\it Journal of Chemometrics}, 5(3), 227-239, 1991.}

\bibitem{TJELMELAND}
{TJELMELAND, H.; LUND, K. V. Bayesian modelling of spatial compositional data. {\it Journal of Applied Statistics}, 30, 87-100, 2003.}


\end{thebibliography}
\end{document}